# *SIM PlanetQuest*

## The Most Promising Near-Term Technique to Detect, Find Masses, and Determine Three-dimensional Orbits of Nearby Habitable Planets

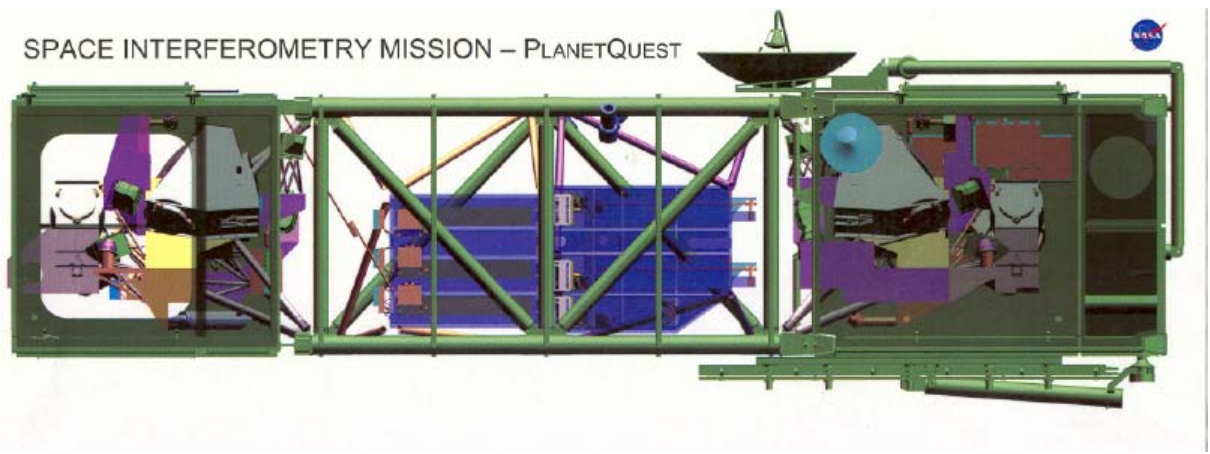

M. Shao (JPL), G. Marcy (Univ. California), S. Unwin (JPL), R. Allen (STScI), C. Beichman (MSC/Caltech), J. Catanzarite (JPL), B. Chaboyer (Dartmouth), D. Ciardi (MSC/Caltech), S. J. Edberg (JPL), D. Gallagher (JPL), A. Gould (Ohio State), T. Henry (Georgia State), K. Johnston (USNO), S. Kulkarni (Caltech), N. Law(Caltech), S. Majewski (Univ. Virginia), J. Marr (JPL), N. Law (JPL), X. Pan (JPL), S. Shaklan (JPL), E. Shaya (Univ. Maryland), A. Tanner (JPL), J. Tomsick (Univ. California),  A. Wehrle (Space Sci Inst), G. Worthey (Washington State Univ)

# *SIM PlanetQuest*
## The Most Promising Near-Term Technique to Detect, Find Masses, and Determine Three-dimensional Orbits of Nearby Habitable Planets

**Introduction**

The past two *Decadal Surveys in Astronomy and Astrophysics* recommended the completion of a space-based interferometry mission, known today as SIM PlanetQuest, for its unique ability to detect and characterize nearby rocky planets (Bahcall 1991, McKee & Taylor 2001), as well as contributions to a broad range of problems in astrophysics. Numerous committees of the National Research Council as well as NASA Roadmaps have similarly highlighted SIM as the one technology that offers detection and characterization of rocky planets around nearby stars and which is technically ready.

*To date, SIM remains the only program with the capability of detecting and confirming rocky planets in the habitable zones of nearby solar-type stars.* Moreover, SIM measures masses and three-dimensional orbits of habitable planets around nearby stars (within 25 pc); these are the only stars for which follow-up by other techniques is feasible, such as space-based spectroscopy, ground-based interferometry, and of course TPF.

**1. Astrometric Detection of Planets**

SIM is an optical Michelson interferometer with a 9-meter baseline. Fringe detection with a CCD allows measurement of wavefront delays, leading to an angular precision of 1.0 µas in a single measurement and a systematic error floor below 0.1 µas (for longer integration times) as verified with the SIM optical testbed at JPL. Indeed, in 2001 NASA HQ established eight technology gates for SIM, and all eight gates have been met and verified (see Section 2 and Marr 2007).

SIM measures angular positions of stars to within 1.0 µas for stars as faint as $10^{th}$ mag (in V) relative to $10^{th}$ mag reference stars within ~1 degree, in ~20 minutes.

The planet detection thresholds by SIM follow directly. As a benchmark, an Earth-mass planet orbiting 1 AU from a solar mass star at 6 pc induces an angular wobble with a radius of 0.5 µas. SIM can detect this since its demonstrated systematic error floor is below 0.1 µas With N measurements spanning a few orbits, the detection improves as the square root of N because the signal's periodicity is coherent (inconsistent with random or systematic errors).

While radial velocity (RV) detection is more sensitive to close-in orbits, astrometry is more sensitive for more distant orbits (Fig. 1). Around stars more luminous than the Sun, an Earth in the habitable zone (HZ) can be detected by SIM in a large volume of space in the solar neighborhood.

*SIM could detect Earth-mass planets orbiting at mid-habitable zone around every one of the most favorable ~64 stars in the solar neighborhood (see Table 1 & Sec. 3a.)*

Many of the terrestrial planets found by SIM can be followed up by ambitious observations made with telescopes on the ground and in space. Moreover, SIM's direct measurement of the astrometric wobble yields the mass of the planet unambiguously (no sin *i* ambiguity). Equally important, the full three-dimensional orbit is derived from SIM astrometry, yielding the orbital eccentricity of single planets and the coplanarity of multiple planets in a system. Coplanarity is often assumed, but without any observational evidence; SIM measures it unequivocally.

*SIM offers the unique ability among proven techniques to both detect the habitable planets around the nearest stars and to give the masses, eccentricities and coplanarities that are so important for testing planet-formation theory.*

With the technology fully tested and vetted, SIM is ready to become the first mission to identify and analyze rocky planets orbiting stars in the solar neighborhood. Such detections would motivate numerous follow-up observations especially from space, launching the new field of habitable planet science. The arguments for SIM were appreciated fully by the two Decadal Surveys that scrutinized it and strongly endorsed it.

**Table 1. SIM Habitable Planet Survey**

| Mass sensitivity at mid HZ | $1 M_\oplus$ | $2 M_\oplus$ | $3 M_\oplus$ |
|---|---|---|---|
| # of target stars | 64 | 149 | 242 |
| # Terrestrial planets (0.3 - 10 $M_\oplus$) | 22 | 34 | 42 |
| # Terrestrial planets in the HZ | 5 | 8 | 9 |
| Total number of planets | 36 | 64 | 90 |

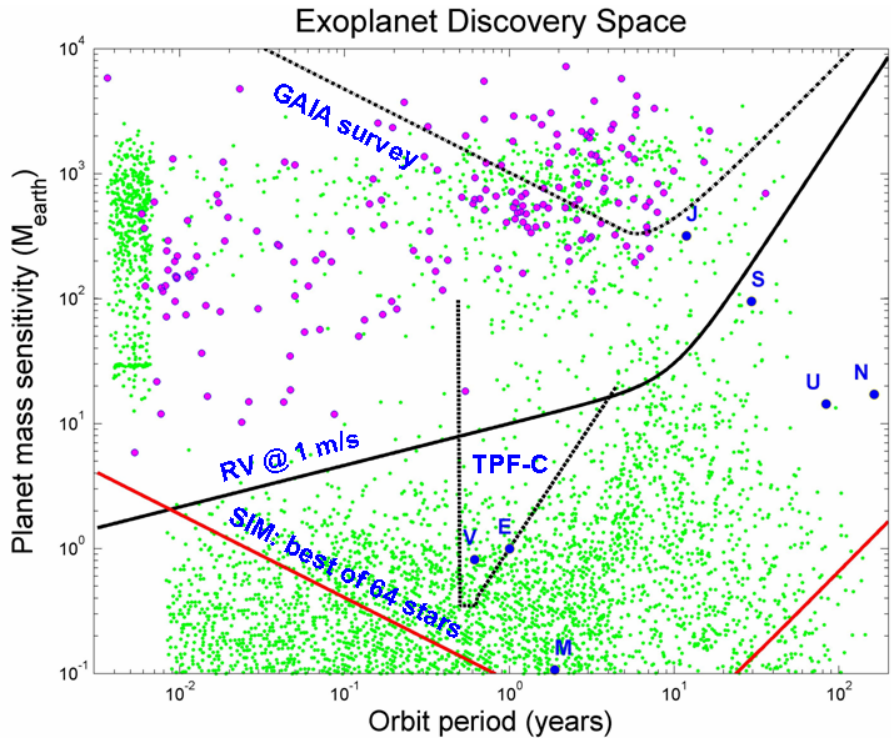

**Figure 1:** SIM is the only mission that explores almost the entire range of masses and orbit periods where terrestrial planets form and evolve. The red curve is for the most favorable of 64 candidate stars, while the least favorable star is about an order of magnitude above this line. (The discovery space lies above each line). Green dots represent planet distributions from the simulations of Ida & Lin (2004ab), and purple dots show the known RV detections.

## 2. SIM is Ready to Be Built

The SIM technology development program was completed to TRL (Technical

Readiness Level) 6 in June 2005. It has been extensively reviewed and was formally signed off by NASA HQ in March 2006. The technology program has demonstrated the extreme performance (1.0 µas narrow-angle single-measurement and 4.0 µas wide-angle mission accuracy) needed for exoplanet detection with substantial margins. Instrumental errors in the SIM testbed (chopped) integrate down as $1/\sqrt{T}$ to a systematic error floor of less than 100 nano-arcsec (Fig. 2).

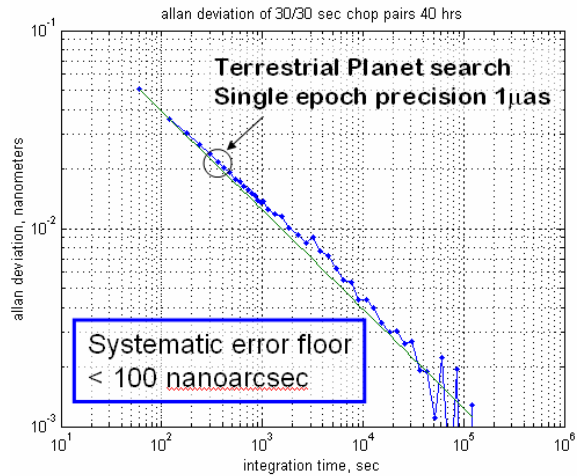

**Figure 2: SIM testbed demonstration of differential (chopping) measurements. The instrument floor lies below 0.1 µas.**

SIM has completed virtually all of its Formulation Phase (A/B) work, including completion of technology development as described above. This has resulted in a mature design with well understood cost and schedule (also verified by NASA external independent reviews).

SIM Implementation Phase (C/D) cost, including launch vehicle, in FY2005 dollars is $1,200M. This estimate is extraordinarily robust, with: 54% budget reserve on the instrument; 43% budget reserve on project cost excluding launch vehicle; and 19% budget reserve on the launch vehicle. The plan also contains 20% schedule reserve.

Operations (Phase E) seven year cost is $365M (FY05$), including four month science verification, five year science operations, and 20 month archive/closeout, representing a cost of roughly $60M/year during operations, including funding for science investigations.

The United States has invested over $500M so far in maturing SIM and its needed technology and reducing implementation risk. *SIM is ready to go now*, providing taxpayers with the payoff – the discoveries that await us in the uncharted realm of extra-solar terrestrial planets.

**3. SIM Exoplanet Search Programs**

The known exoplanets have orbital period and mass distributions that are roughly uniform in log-period and log-mass, respectively. Analysis of 200 known planets yields power laws for the distribution in mass and period of $dN/dM \sim M^{-1}$ and $dN/dP \sim P^{-1}$ (Cumming et al. 2007).

SIM detects the presence of a planet by measuring the periodic reflex motion of the star as the planet orbits the star. Ideally, the motion is observed for at least one orbital period. If the orbital period is longer than the duration of observations, planet detectability is degraded due to confusion with proper motion. After a series of observations has been made, we may compute the periodogram power with the detection threshold set so that the false alarm probability is less than 1%.

The SIM exoplanet program consists of three major components.

**(a) A habitable planet survey** focusing on the nearest stars. These are planets between 0.3 and 10 times the mass of the Earth, orbiting in the habitable zone, which ranges from 0.7 to 1.4 AU from the star, scaled appropriately for stellar luminosity.
**(b) A broad planet survey** of ~2100 stars. The broad survey at ~4 times lower

astrometric accuracy but with ~20 times as many targets will be able to find terrestrial mass planets in five year orbits around low mass stars.

**(c) A young star planet survey** focusing on Jupiter/Saturn mass planets orbiting very young stars, some of which are young enough that the planet formation and migration processes are still underway.

Section 4 describes how SIM is synergistic with other planet detection techniques.

### 3a. SIM Habitable Planet Survey

The SIM habitable planet survey can be optimized by considering fiducial AFGK stars in the solar neighborhood. One may calculate the orbital distance to the HZ for each star and rank the stars in terms of astrometric detectability.

For the habitable planet survey, each target is allocated just enough observing time to detect a one Earth mass planet at the middle of its habitable zone. Using 37% of SIM time in a five year mission, SIM can conduct a HZ search of 64 stars with sensitivity down to one Earth mass. Using 37% of a ten year mission, SIM could search 98 of the nearest stars down to one Earth mass in the HZ. The amount of observing time needed increases as $1/(\text{planet mass})^2$ so that a larger number of stars can be searched by relaxing the mass sensitivity to two or three Earth masses. (See Table 1.)

The science return can be estimated based on the expected distribution of planets in mass and orbital parameter space. From the 200 known exoplanets spanning the range from 7 Earth masses to 10 Jupiter masses and orbital radii from 0.05 AU to 3 AU, the probability distributions of mass and orbit radius can be extrapolated into the long-period and low-mass regime. Recent theoretical models of rocky planet formation (Ida & Lin 2004ab, 2005 and Benz et al. 2006) predict that there are many more <10 $M_\oplus$ planets than Jupiters. Benz expects a ~5 fold increase in terrestrial planets over a simple extrapolation of the power-law model for higher mass planets.

Accordingly, we adopt for our model of the probability distribution of planets the power laws derived from the known RV-detected planets by Cumming et al. (2007), but bump up the density of planets below 10 $M_\oplus$ by a factor of five. This model predicts that ~73% of stars have a terrestrial planet, ~10% of stars have terrestrial planets in the HZ and ~1% of stars will have Earth 'clones' (0.8~1.2 $M_\oplus$, 0.7~1.4 AU). Using this model one may estimate how many terrestrial planets SIM would find, as indicated in Table 1. The model also predicts that ~22% of stars have planets more massive than ten Earth masses.

### 3b. SIM Broad Planet Survey

The SIM broad planet survey is a comprehensive search for planets, down to terrestrial mass, around ~2100 stars. These target stars cover essentially the full range of stellar properties and the hundreds of planets found will form the largest sample of nearby planets with precisely known orbits and masses. This unique sample will set the target lists for future exoplanet imaging and spectroscopic missions.

The survey, comprising just 4% of SIM mission time, has been designed to detect terrestrial-mass and larger planets with orbital periods of several years. To enable coverage of a large number of targets, the survey has a somewhat lower astrometric precision than the HZ planet-search portion of the SIM mission.

This survey has a nominal single-measurement astrometric precision of 4 μas, which allows unambiguous detection of 2-Earth-mass planets in five-year orbital periods around solar-mass stars at 10 pc distance (1% false-alarm probability). The survey is sensitive to sub-Earth-mass planets around M-dwarfs and terrestrial-mass planets around all but the most massive stars in the sample. Multiple-planet systems will be readily identifiable. Precise three-dimensional orbital solutions will provide mass, orbit semimajor axis, inclination, and eccentricity for each detected planet and test the ubiquity of planetary system coplanarity. The survey will detect planets in all orbital orientations, allowing unambiguous statements to be made about the existence or absence of planets around individual stars, unlike transit and RV surveys.

**Table 2: SIM Broad Planet Survey**

| Number of target stars | 2100 |
|---|---|
| # Terrestrial planets (0.3 - 10 $M_\oplus$) | 84 |
| # Neptunes (10 - 30 $M_\oplus$) | 42 |
| # Jupiters (30 $M_\oplus$ - 10 $M_{Jupiter}$) | 300 |
| Total number of planets | 426 |

This survey is complementary to the GAIA planet survey. GAIA covers approximately ~50,000 targets but with a much lower (~70 µas) equivalent single-measurement precision. While the GAIA mission will detect Jovian planets, the SIM broad planet survey will detect planets ~20 times less massive.

The survey targets 2,100 stars. The sample is split into several groups, each designed to answer specific questions about the exoplanet population.

The primary target group will provide detailed statistics on the planetary populations around the full range of main-sequence stars. Although exceptional sensitivity will be achieved for every target, particular emphasis will be placed on the very low mass, high mass, and active stars that are extremely difficult to search with other planet-detection techniques.

Other target classes include stars in a wide range of age and metallicity, stars with dust disks, white dwarfs, stars in multiple systems, and stars with known exoplanets.

Since SIM is a pointing observatory, other interesting targets will be searched for low-mass planets as they are discovered.

We emphasize that the broad survey will produce a complete planetary census of the 2,100 target stars down to the terrestrial-mass regime. Planet formation simulations (Ida & Lin 2004 ab, 2005; Benz et al. 2006) as well as the microlensing detection of a rocky exoplanet (Dominik et al. 2006) suggest that terrestrial-mass planets are far more common than gas giants.

### 3c. SIM Young Star Survey

SIM's Young Stars and Planets Key Project will contribute a census of gas giant planets orbiting 150-200 stars with ages from 1 Myr to 100 Myr. These discoveries will lead to greater understanding of the formation and dynamical evolution of gas giant planets. The groundbreaking aspect of this investigation is not its mass sensitivity but its unique ability to probe an important epoch in the planet formation process about which little is currently known.

Ground-based advances in detecting young Jupiter-mass planets are likely to come from IR coronagraphic imaging and extreme AO. Such techniques have, to date, produced a few intriguing objects at wide separations (20-100 AU) from very young (<10 Myr) host stars, e.g. 2M1207 (Chauvin et al. 2005) and GQ Lup (Neuhauser et al. 2005). These companions potentially have masses well into the planetary regime, depending on the choice of evolutionary models that predict the brightness of "young Jupiters" as a function of mass and age (Baraffe et al. 2002). Unfortunately, dynamical determinations of mass (which would distinguish planets from brown dwarfs) are impossible for companions on such distant orbits. Multiple fragmentation events (Boss 2001), rather than core accretion in a dense disk (Ida & Lin 2004) may be responsible for the formation of these distant objects.

A Jupiter orbiting ~1 AU away from a 0.8 $M_\odot$ star at the distance of the youngest stellar associations (1-10 Myr), such as Taurus and Chamaeleon, would produce an astrometric amplitude of 8 µas, easily detected with SIM. At the 25-50 pc distances of the nearest young stars (10-50 Myr), such as members of the β Pic and TW Hya groups, the same stellar system would have an enormous astrometric amplitude in excess of 100 µas.

In a survey of 200 young stars, SIM can expect upwards of 10-20 planetary systems (assuming that only the presently known fraction of stars, 5-10%, has planets). The sensitivity threshold has been set to ensure the detection of Jupiter-mass planets in the critical orbital range of 1 to 5 AU. These observations, when combined with the results of planetary searches of mature stars, will allow us to test theories of planetary formation and early solar system evolution. By searching for planets around pre-main sequence stars carefully selected to span an age range from 1 to 100 Myr, we will learn at what epoch and with what frequency giant planets are found at the water-ice "snowline" where they are expected to form (Pollack et al. 1996). This will provide insight into the physical mechanisms by which planets form and migrate from their place of birth, and about their survival rate.

These observations will provide data, for the first time, on such important questions as: What processes affect the formation and dynamical evolution of planets? When and where do planets form? What is the initial mass distribution of planetary systems around young stars? How might planets be destroyed? What is the origin of the eccentricity of planetary orbits? What is the origin of the apparent dearth of companion objects with masses between gas giant planets and brown dwarfs seen in mature stars? How might the formation and migration of gas giant planets affect the formation of terrestrial planets?

About half of the sample will be used to address the "where" and "when" of planet formation by observing classical T Tauri stars which have massive accretion disks as well as weak-lined T Tauri stars. The other half of the sample will be stars from 5 Myr, through 10 Myr (thought to mark the end of prominent disks), and ending around the 100 Myr age, to study the effects of dynamical evolution and planet destruction (Lin et al. 2000).

## 4. SIM Astrometry Complements Other Exoplanet Search and Characterization Techniques

Astrometry and radial velocity are complementary approaches to planet detection. RV is more sensitive to short period planets and astrometry more sensitive to long period planets. The SIM science team has members from the two leading RV planet search groups. Extensive RV data from current instruments have been studied to derive the sensitivity limits due to motion of the stellar atmospheres of the target stars. At ~1m/sec it is temporally correlated stellar noise, caused by p-modes, super granulation, meridional oscillations, and magnetic cycles. These stellar velocity fields have characteristic time scales of days to decades, causing temporally correlated Doppler shifts of 2 m/s (for G stars) and 1 m/s (for K stars). This 1 m/s correlated noise floor prevents the Doppler method from securely detecting planets under 10 $M_\oplus$ beyond 0.5 AU from the star.

Astrometry is the only technique that permits accurate measurement of both mass (a key parameter for planetary science) and orbit inclination. The RV technique has revealed 20 multi-planet systems, 13% of the total number of known exoplanets.

However we do not know if coplanar orbits are common or rare. SIM astrometry can answer this key question. SIM can also detect Saturn-mass planets in five year orbits around solar-like stars at 800 pc. Finding additional planets in large orbits around planets found by missions such as Kepler will tell us if our own solar system is common. The completeness of astrometry is important in that non-detection is significant.

The SIM astrometric search is also useful as a precursor to planning a direct detection campaign, such as with TPF. Terrestrial planets in the HZ that are found by SIM would become prime targets for a follow-up space mission to obtain spectra of the planets: both missions study samples of nearby solar-type stars. SIM astrometry also provides information about where and when to observe with TPF, i.e. avoiding TPF observations when the planet resides within the inner working angle. Non-detection by SIM is also important, helping to prioritize TPF target stars.

## 5. Other Astrophysics with SIM

Precision mass and orbital measurements of *stellar companions* by SIM will be made across the whole H-R diagram and through the metallicity dimension, tying down the mass-luminosity relation, from the hottest O stars through the coolest M stars, and into the brown dwarf regime.

The study of dark matter is the goal of several SIM key projects. The revolution of stars in the galactic disk and the motion of tidally disrupted dwarf galaxies in the halo of the Milky Way will enable astronomers to map the distribution of dark matter in our galaxy. Measurements of gravitational microlensing by SIM, made jointly with ground-based observations, will yield the Galaxy's mass spectrum inside the Sun's galactic orbit. Mass determinations of dark and faint objects including planets, brown dwarfs, low mass stars, white dwarfs, neutron stars, and black holes will be made with this technique. SIM will measure the proper motion of ~20 galaxies in the local group. These Key Projects and others already selected (by NASA AO) will be making a variety of other astrophysics investigations involving extreme objects, ages of star clusters, and the mechanism of quasar emission. Approximately 40% of SIM science time is unallocated, and will be made available through open competition.

## 6. SIM Summary

SIM is capable of detecting habitable planets around ~100 nearby stars, and will provide a census of planets around ~2100 stars to place our solar system in a broader context. SIM will open a new era of the characterization of rocky planets in the solar neighborhood.

Having completed its technology program in 2005, and almost completed Formulation Phase (Phase B), SIM is a low-risk mission with a well-established cost. It could launch as soon as 2015. Humanity could discover its first one Earth mass planet in the HZ less than a decade from now!

## 7. References

Bahcall, J.N. 1991, "The Decade of Discovery in Astronomy and Astrophysics, National Academies Press, ISBN 0-0309-04381-6.
Baraffe, I. et al. 2002 A&A, 382, 563
Benz, W. et al. 2006 Proceedings IAU Colloquium 200
Boss, A. 2001 ApJ, 551, L167
Chauvin, G. et al. 2005, A&A, 438, L25
Cumming, A., et al., 2007 ApJ, in press
Dominik, M. et al., 2006 A&G 47, 3, 25
Ida, S. & Lin, D.N.C 2004 ApJ, 604, 388
Ida, S. & Lin, D.N.C 2004 ApJ 616, 567
Ida, S. & Lin, D.N.C 2005 ApJ 626, 1045
Lin, D.N.C. et al. 2000, Protostars and Planets IV, 1111
Marr, J. C. et al. 2007, ExoPTF White Paper, Technology for Astrometric Detection of Nearby Earth-Mass Habitable-Zone Planets from Space
McKee, C. & Taylor, J., 2001 Astronomy and Astrophysics in the New Millennium, National Academies Press, ISBN–309-0731-2
Neuhäuser, R. et al. 2005, A&A, 435, L13
Pollack, J., et al. 1996, Icarus 124, 62